\documentstyle[aps,multicol,prl,epsfig]{revtex}
\begin{document}
\begin{title}
{\bf Focusing Revisited: an MN-dynamics Approach}
\end{title}

\author{C.I. Siettos and I.G. Kevrekidis}
\address{Department of Chemical Engineering, Princeton University,
6 Olden Str. Princeton, NJ 08544}
\author{P.G. Kevrekidis}
\address{
Department of Mathematics and Statistics, University of Massachusetts,
Amherst, MA 01003-4515, USA}
\date{\today}
\maketitle

\begin{abstract}
The nonlinear Schr{\"o}dinger (NLS) equation is a ubiquitous example of
an envelope wave equation for conservative, dispersive systems.
We revisit here the problem of self-similar focusing of waves in the case
of the focusing NLS equation through the prism of a dynamic renormalization
technique (MN dynamics) that factors out self-similarity
and yields a bifurcation view of the onset of focusing.
As a result, identifying the focusing self-similar solution becomes
a steady state problem.
The discretized steady states are subsequently obtained and
their linear stability is numerically examined.
The calculations are performed in the
setting of variable index of refraction, in which the onset of
focusing appears as a supercritical bifurcation of a novel
type of mixed Hamiltonian-dissipative dynamical system
(reminiscent, to some extent, of a pitchfork bifurcation).
\end{abstract}

\vspace{3mm}

Self-focusing and wave collapse have received a considerable
amount of attention in many diverse branches of physics, ranging
from optics \cite{Sulem,Rasmussen} to
hydrodynamics \cite{Baren}
and from gravity \cite{Chopt} to plasma physics \cite{Ber}.
The interested reader
can also consult \cite{Gold,Brenner}.

The  most frequently studied model equation in the context
of self-focusing and the ensuing blow-up of the relevant
solutions is the two-dimensional (2D) and three-dimensional
(3D) nonlinear Schr{\"o}dinger (NLS) equation.
This benchmark example is
well-known (see e.g., \cite{Sulem} and references therein) to exhibit
unstable solitary wave solutions which collapse in finite time.
Numerous publications have addressed
the focusing properties of the solutions (Refs. \cite{Pap,Zakh}
can only claim to give a number of representative examples of this
large volume of work).
Recently, a revived interest in the subject
has led to numerous new insights \cite{Budd}.

Here we take a slightly different approach to the same problem, viewing it
from the perspective of dynamical systems theory and making the
connections to both earlier work on the NLS problem \cite{Pap,Zakh}
as well as more recent work on self-similar blowup
problems \cite{Betelu}.
Many of the results that are
presented here on NLS were previously known from the works
of \cite{Pap,Zakh}.
Such results include the effective factoring out of the self-similarity
to view focusing as a steady state problem, as well as the continuation
of the branch of focusing solutions.
   What we believe is new in our exposition is the intuition it
provides by identifying the onset of focusing  as a steady state bifurcation
of a (transformed) partial differential-algebraic equation (PDAE).
This appears to us to be a new type of bifurcation
reminiscent of a supercritical pitchfork.
The identification of focusing solutions as steady states
of this system also allows us to linearize around them and examine
(as well as justify) their stability and the ensuing structure of the
bifurcation diagram. 
Here these computer-assisted tasks are performed for a finite 
discretization of the problem.

Finally, it is worth noting that this approach has been implemented in a somewhat different but
equivalent form of the regular blow-up NLS problem.
In particular, in previous studies see e.g.,
\cite{Sulem,Pap,Zakh}, the radial form of the NLS equation was
analyzed,
\begin{eqnarray}
R_{rr} + \frac{d-1}{r} R_r-R+|R|^2 R=0,
\label{cseq1}
\end{eqnarray}
where $R=R(r)$ denotes the radial part of the solution.
The analysis (when performed parametrically) would then involve
a slightly unphysical ``bifurcation parameter'', namely the spatial
dimensionality $d$.
In contrast to that, but equivalently for the purposes of the analysis, we use
a different (but more physically realistic) bifurcation parameter.
The latter is the power of the nonlinearity $\sigma$ in the model:
\begin{eqnarray}
i u_t = - \Delta u - |u|^{2 \sigma} u;
\label{cseq2}
\end{eqnarray}
$u$ is the complex envelope of the electric field (in optical applications)
and the present setting arises when the index of refraction has a power
law dependence ($\sigma$ being the relevant power) on the intensity
of the field.
It is well known that the model of Eq. (\ref{cseq2})
becomes critical (i.e., unstable to wave collapse) for $\sigma \geq 2/d$,
where $d$ is the spatial dimensionality of the Laplacian in (\ref{cseq2})
(see e.g., \cite{Sulem,Wein}).
In view of that, we choose to study
the model in {\it one spatial dimension}, using $\sigma$ as a bifurcation
parameter, with $\sigma_{cr}=2$ in that case. Notice that for $\sigma=1$,
the model becomes the celebrated integrable example of the 1D cubic NLS
\cite{Abl}.

When considering self-focusing solutions of Eq. (\ref{cseq2}) it is
customary to make a straightforward appropriate scaling.
In particular, using
$u=L^{-1/\sigma} v(\xi,\tau)$, with $\xi = x/L$, $\tau=\int_0^t L^{-2}(t) dt$,
one  obtains
\begin{eqnarray}
i v_{\tau}=-v_{\xi \xi}-|v|^{2 \sigma} v-
i a(\tau) (\frac{1}{\sigma} v + \xi v_{\xi}),
\label{cseq3}
\end{eqnarray}
where $a(\tau)=-L_{\tau}/L$ (see e.g., \cite{Sulem}).
A direct pointer to the criticality of $\sigma=2$ is the fact that
the norm is preserved under the rescaling only for that value of $\sigma$.
Hence, dynamical (i.e., permitted by the time evolution) rescaling
for the original NLS model of Eq. (\ref{cseq2}) is {\it only} 
permissible for $\sigma=2$.
On the other hand,  Eq. (\ref{cseq3}) has appeared
in various forms in a number of works on
the focusing problem such as \cite{Pap,Zakh} (see \cite{Sulem} for a
review of the relevant methods and results), in the context of what
was called ``dynamic rescaling''.
The idea behind such
a rescaling is that the scaling
is chosen in such a way that the rescaled time goes to infinity
as the (finite) time of the singularity $t^{\star}$ is approached.
The closure of the system (with an equation for $a(\tau)$) was given
by imposing an arbitrary, but physically or numerically motivated,
constraint, such as the preservation of a certain norm of the solution.

Here, we will follow a slightly different path, motivated from a general
approach to self-similar problems recently presented in \cite{Betelu}.
In particular, if we consider the equation
\begin{eqnarray}
i u_t= F(u),
\label{cseq4}
\end{eqnarray}
typically the potential for self-similar solutions arises when a
scaling is inherent in the nonlinear operator of the right hand side
(RHS) of (\ref{cseq4}).
In general:
\begin{eqnarray}
F \left(B u(\frac{x}{L}) \right)= L^c B^d F \left( u(\xi) \right),
\label{cseq5}
\end{eqnarray}
where $\xi=x/L$ (keeping in line with the notation used in earlier
works such as \cite{Sulem,Pap}).
The dispersive operators with on-site nonlinearity considered here
are however different from the ones of \cite{Betelu} in that the
presence of the onsite potential imposes a certain restriction
on the scaling through Eq. (\ref{cseq5}), namely for the $F$ of
the RHS of \ref{cseq4})
\begin{eqnarray}
B &=& L^{-\frac{1}{\sigma}}
\label{cseq6}
\\
F \left(B u(\frac{x}{L}) \right) &=& \frac{1}{L^2} F \left( u(\xi) \right).
\label{cseq7}
\end{eqnarray}
Using in the equation
the ansatz $u=B(\tau) v(x/L(\tau),\tau)$, where $\tau$ is an,
as of yet, undetermined function of the original time $t$, we obtain:
\begin{eqnarray}
i \left( \frac{B_{\tau}}{B} v - \frac{L_{\tau}}{L} \xi v_{\xi}
+  v_{\tau} \right) \tau_t = -\frac{1}{L^2} \left( v_{\xi \xi} +
|v|^{2 \sigma} v
\right),
\label{cseq8}
\end{eqnarray}
where the subscripts denote respective partial derivatives.
Notice also that
using Eq. (\ref{cseq6}), $B_{\tau}/B=-L_{\tau}/(\sigma L)$ and hence the
self-similarity has a single scaling factor, $L$.
Time parametrization
can be chosen on the basis of convenience and hence we use here
$\tau$ such that $\tau_t=1/L^2$.
If we look, as is customary (see e.g.,
\cite{Sulem} and references therein) for a {\it standing wave} solution of
Eq. (\ref{cseq8}), we use the so-called phase invariance to look for
solutions of the form $v \rightarrow \exp(i \Lambda \tau) v$, which gives
the same equation as before but with an extra $\Lambda v/L^2$
in the RHS.
Without loss of generality, we set $\Lambda=1$.
Finally, one has the one-parameter family
($L$) freedom of the group orbit of self-similarity
(see \cite{Rowley} for the case of traveling solutions; the case of
self-similarity
is treated in \cite{Betelu}).
To determine the scaling factor, we lift the symmetry-induced degeneracy
through a pinning condition \cite{Taylor} of the form
\begin{eqnarray}
\int_{-\infty}^{\infty} Re(v(\xi,\tau)) T(\xi) d \xi= C,
\label{cseq9}
\end{eqnarray}
where $C$ is a constant.
$T(\xi)$ is an arbitrary, so-called ``template'' function.
Alternatively, one can consider maximazing the inner product with
the template (such an approach has been considered in \cite{Rowley}
for settings with travelling wave solutions, as well as ones with
self-similar solutions).
Differentiating Eq. (\ref{cseq9}) and using eq. (\ref{cseq8}), we arrive
to the following equations for the real and imaginary parts of the
solution $v=U+ i W$
\begin{eqnarray}
U_{\tau} &=& -W_{\xi \xi}-(U^2+W^2)^{\sigma} W + W - G
(\frac{1}{\sigma} U + \xi U_{\xi})
\label{cseq10}
\\
W_{\tau} &=& U_{\xi \xi}+ (U^2+W^2)^{\sigma} U - U - G
(\frac{1}{\sigma} W + \xi W_{\xi}),
\label{cseq11}
\end{eqnarray}
where
\begin{eqnarray}
G=-\frac{L_{\tau}}{L}= - \frac{\int_{-\infty}^{\infty}
\left[ W_{\xi \xi} + (U^2+W^2)^{\sigma} W - W \right] T(\xi) d \xi}{
\frac{C}{\sigma} + \int_{-\infty}^{\infty} \xi U_{\xi} T(\xi)
d \xi}.
\label{cseq12}
\end{eqnarray}
Eqs. (\ref{cseq10})-(\ref{cseq11}) supplemented with Eq. (\ref{cseq12})
constitute the MN-dynamics formulation of the focusing problem for the
NLS equation.

\begin{figure}[h]
\centerline{\psfig{file=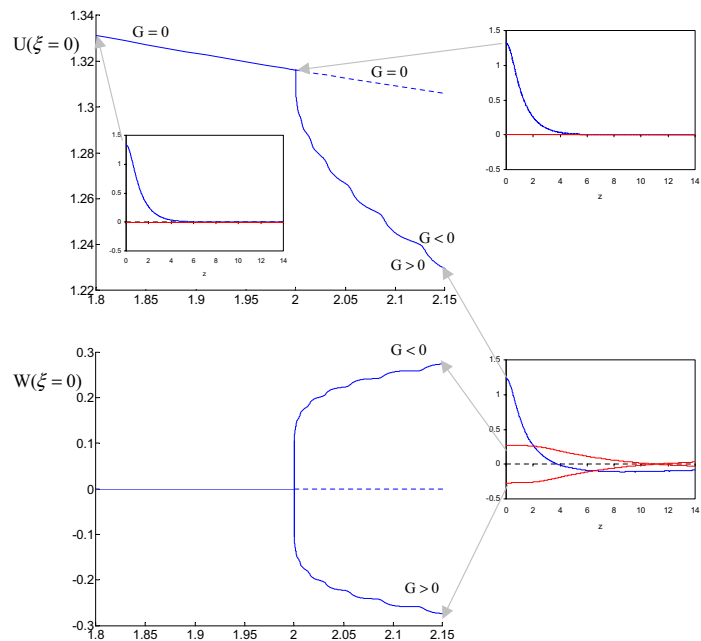,height=5in}}
 \caption{The figure shows the variation as a function of
$\sigma$ of the real (top panel) and imaginary (bottom panel)
parts of the field evaluated at  $\xi=0$.
The soliton branch is stable for $\sigma<2$ (solid line), while
it becomes unstable (dashed line) for $\sigma \geq 2$, giving
rise to the blowup branches (with $G \neq 0$). The insets show 
the spatial profile of the solution for the positive semi-axis.}
\label{csfig1}
\end{figure}

The regular solitary wave solutions of the equation (\ref{cseq2}) exist
as stationary solutions of the equation ensuing from (\ref{cseq2}) upon
the substitution $u= \exp(i t) v$, or equivalently as ``trivial'' focusing
solutions of the MN equations (\ref{cseq10})-(\ref{cseq11}) with a zero
blowup rate, $G=0$.
These solutions are stable for $\sigma < 2$.
Their
mechanism of instability involves a pair of linearization eigenvalues
that bifurcate from the band edge of the continuous spectrum, for
$\sigma > 1$.
The dispersion relation describing the continuous spectrum
is $\lambda= \pm i (1 + k^2)$, where $\lambda$ is the eigenvalue corresponding
to wavenumber $k$, for our case of $\Lambda=1$.
This pair of eigenvalues
moves towards the origin of the spectral plane, where 2 more pairs of
eigenvalues lie due to the 2 invariances (translational and phase) of
Eq. (\ref{cseq2}).
At $\sigma=2$,
the additional pair ``arrives'' at the
origin, thereafter exiting along the real axis, rendering the soliton
branch unstable as is shown in Fig. \ref{csfig1}.

Beyond the instability threshold, the soliton is well-known to be
unstable towards focusing, resulting in a self-similar blowup for
$\sigma > 2$.
Such solutions are most commonly obtained through
integration of the dynamic rescaling equations (or even possibly
of the original equations in an adaptively refined mesh), see
e.g., \cite{Sulem,Pap,Zakh} and references therein.
Here we use an alternative
perspective\footnote{Our work has a somewhat similar flavor to
Phys. Rev. A {\bf 38}, 3837 (1988), but is posed in a different,
we believe slightly more general framework that allows us to perform
a systematic bifurcation analysis of the results.}.
We identify the soliton branch as a special solution
(with $G=0$) of Eqs. (\ref{cseq10})-(\ref{cseq11}).
We then use a ``soliton-motivated" pinning condition.
In particular, we use as our arbitrary
template a Dirac mass, centered at a given point $x_0$, forcing
the real part of the solution to have at $x_0$ the same value as
the soliton does for $\sigma=2$.
Other pinnings (for the imaginary
part, or for combination of the real and the imaginary part) have
also been used with the same results.
%
The pinning condition in
conjunction with Eqs. (\ref{cseq10})-(\ref{cseq11}) allows us to
construct for $\sigma > 2$ the branch of self-focusing solutions
as steady states through the use of Newton iteration and
pseudo-arclength continuation \cite{Doedel}.

\begin{figure}[h]
\centerline{\psfig{file=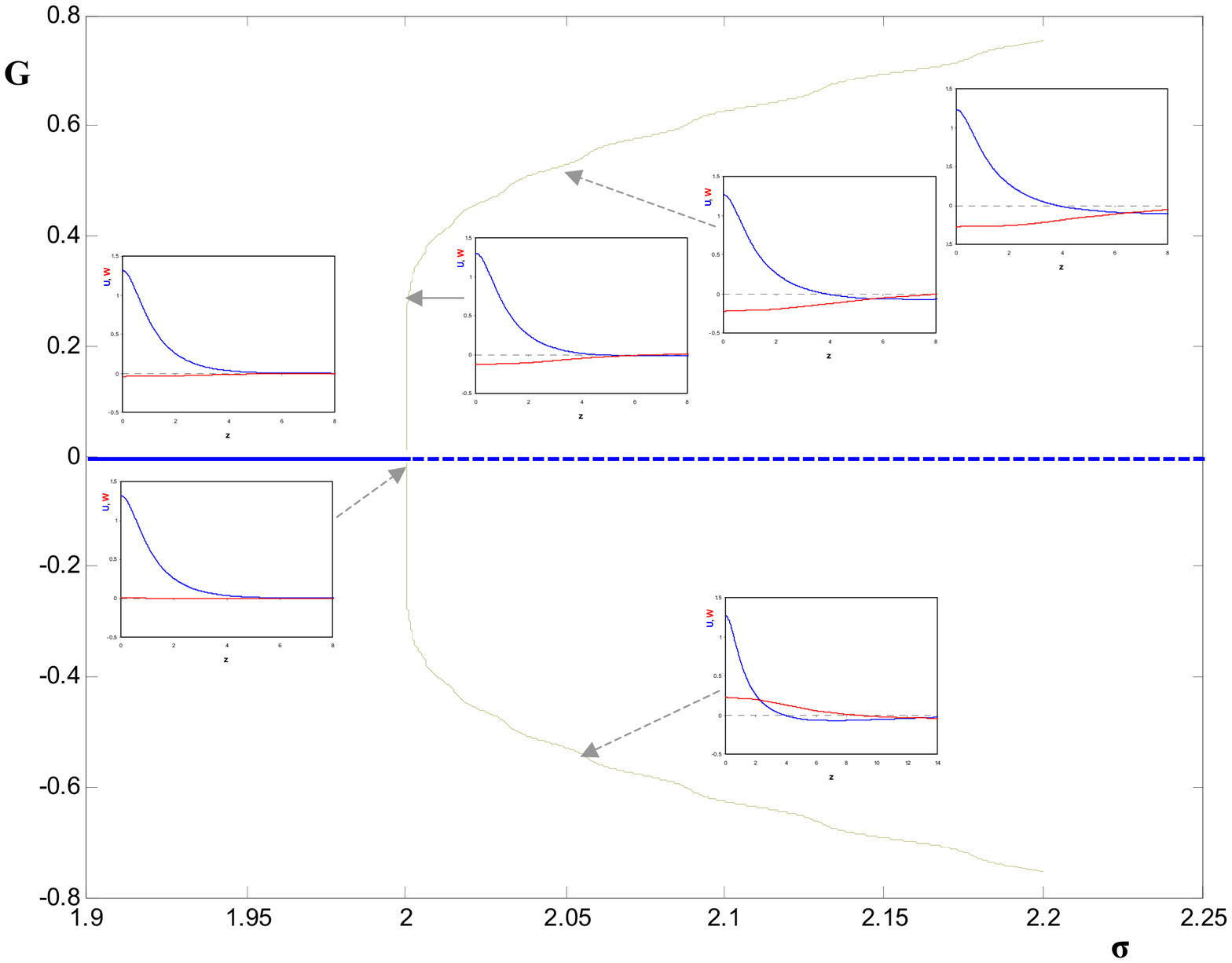,height=4.25in,angle=0}}


\centerline{\psfig{file=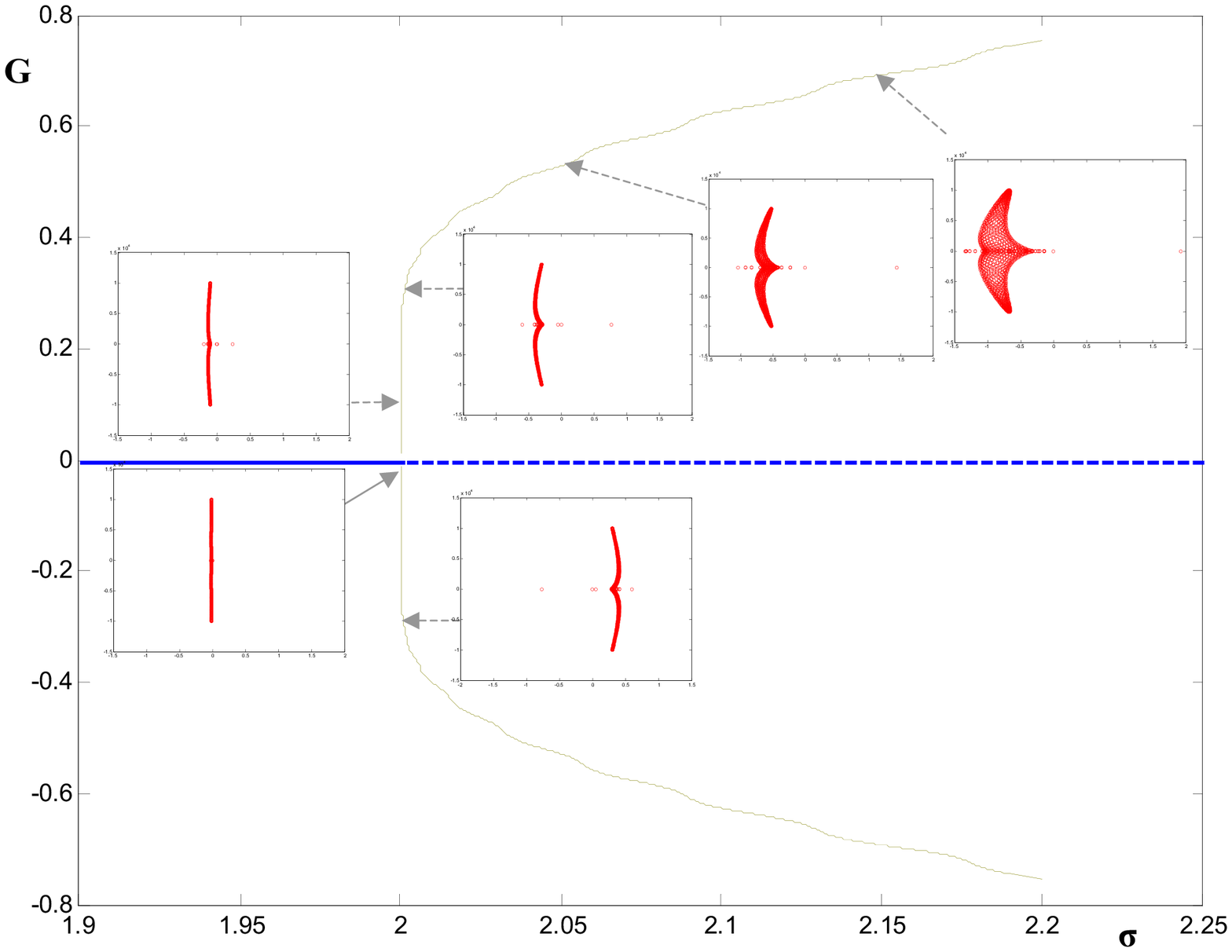,height=4.25in,angle=0}}
 \caption{The top panel shows the MN-dynamics bifurcation
diagram of $G$ vs. $\sigma$. At $\sigma_{cr}=2$, the new branch of
focusing solutions is born. The  panel insets show the profile
of the solution in different points along the branch. The bottom
panel insets show the corresponding MN-dynamics stability analysis
through the linearization eigenvalues. Notice the non-Hamiltonian
nature of the dynamics, as well as the presence of one unstable
(real and positive) eigenvalue in the upper branch (for details
see text). The spectrum of the lower branch is the mirror
symmetric of the upper one with respect to the imaginary axis.
}
\label{csfig2}
\end{figure}

The algebraic problem is solved using the boundary value solver
of the package gPROMS\footnote{see e.g., www.psenterprise.com},
a commercial simulation tool for solving systems of ordinary,
and/or partial differential equations and/or of partial differential-algebraic
equations of index 1.
The pseudo-arclength continuation method was incorporated
within gPROMS.
The computational domain was chosen to be the right half-axis $\xi=0$
(due to symmetry of the solutions sought).
The length and the tessellation of the domain were
chosen appropriately to ensure reliability of the solutions, i.e.,
robustness against further discretization refinement and domain
enlargement.
The solutions were approximated using a centered finite
difference method of order two in a domain of length equal to 14
and a total of 700 discretization intervals, while the absolute
accuracy was set to $1E-06$.
The solutions we obtained are shown in Fig. \ref{csfig2}.

An additional advantage of effectively ``factoring out'' the self-similarity
through  probing the solutions as steady states of the MN problem, stems
from our ability to perform linear stability analysis computations for
these solutions. It should be noted that linear stability within a 
renormalization group approach to self-similar solutions was also considered
in gravitational problems, see e.g., \cite{Chopt}.
Such numerical computations for the truncated problem are shown in
Fig. \ref{csfig2}.
These
computations are for $G > 0$, and clearly indicate that the dynamical
system of Eqs. (\ref{cseq10})-(\ref{cseq11}) is a dissipative
(a non-Hamiltonian) one.
This is also structurally clear in the equations, as terms of the form
$- i v$ are well-known dissipative perturbations of NLS.
Analyzing
the structure of the ``umbrella-shaped'' spectrum in comparison with
the linearization spectrum of the Hamiltonian problem for $G=0$, we
recognize that the genuinely complex eigenvalues constitute the
continuous spectrum.
In addition, we (typically) obtain $3$ eigenvalues along
the real line.
In the Hamiltonian untransformed problem, there are 6 eigenvalues
that are near the origin for $\sigma$ close to 2, 4 of which are
at the origin (two with a corresponding symmetric, zero node eigenvector,
related to phase invariance and two with an antisymmetric, one node
eigenvector, corresponding to translational invariance).
The additional
pair that eventually leads to the instability has a corresponding
symmetric eigenvector with two nodes.
In the dissipative MN system,
we are left with only 3 eigenvalues, since 2 out of the 6 are absent
due to considering only half the domain (where only symmetric
eigenfunctions will be present).
An additional eigenmode is absent due
to imposing the pinning condition (decreasing by 1 the degrees of freedom
of the system).
One of the three eigenvalues is at the origin, as the
MN dynamics also possess a conservation law, namely Eq. (\ref{cseq9}).
The phase invariance is, however, explicitly broken here due to
the presence of the term proportional to $G$.
In fact, one can show
that unless one is ``exactly on'' a steady state of
Eqs. (\ref{cseq10})-(\ref{cseq11}), neither the invariance $v \rightarrow
v \exp(i \phi)$ (for $\phi$ an arbitrary constant), nor the corresponding
conservation law $d ||u||_2/dt=0$ (where $||u||_2$ is the $L^2$ norm of
the solution) are respected, for the pinning of Eq. (\ref{cseq9}).
In fact, it can be straighforwardly predicted
within the realm of  (leading order) perturbation theory
that the bifurcation of the phase eigenvalue for $\sigma \rightarrow 2^{+}$
can be approximated by
\begin{eqnarray}
\lambda \sim G \frac{\sigma-2}{2 \sigma}.
\label{cseq13}
\end{eqnarray}
%
This shows that, as is indeed verified in numerical computations, the
bifurcation of the phase eigenvalue is in the real positive semi-axis of the
spectral plane, for $\sigma > 2$.
However, such an instability (which is a genuine one for the non-phase
invariant MN dynamics) will {\it not} be relevant for the original
norm preserving (Hamiltonian) dynamics and hence the
self-focusing solution will be a stable blowup
solution of Eq. (\ref{cseq2}). I.e., the presence of a conservation 
law in the original NLS model does not permit the dynamical development
of this instability, or, equivalently, does not permit the system to
``explore'' this unstable eigendirection.

As $\sigma$ is increased from the critical value, the generation
of the blowup branches can be seen to occur in a form 
reminiscent of a
supercritical pitchfork-like  bifurcation in Fig. \ref{csfig1}.
Notice the steepness of the variation of $G$ in the neighborhood
of the critical point. This very abrupt increase of the value of
$G$, as well as earlier asymptotic arguments seem to support a
dependence beyond all algebraic orders of $G$ on $\sigma-2$
(i.e., an exponential dependence). However, the rigorous 
proof of such an estimate is beyond the scope of the present work
and will be addressed in a future publication.
However, it should be highlighted that this is a rather unusual,
and most probably novel form of a dynamical system, possessing solutions
with as well as without
focusing, the corresponding dynamics however being significantly
modified (from dissipative to Hamiltonian) between the two cases.
Identifying the normal form behavior of such dynamical systems close
to criticality (close to the onset of focusing) is an interesting open problem.
Furthermore, it should be noted that a natural setting for the study 
of the bifurcations of such dynamical systems appears to be the 
framework 
of the relevant differential-algebraic equations (due to the presence 
of the constraint).

Notice that for $\sigma \rightarrow 2^+$, $G \rightarrow 0$ very abruptly.
For $\sigma=2$, there is no solution with $G \neq 0$ \cite{Pap}. Dynamically,
this is manifested as $\lim_{\tau \rightarrow \infty} G(\tau)=0$, the
approach to the limit being logarithmic \cite{Pap}.
It would be interesting to explore
whether 
travelling problems can exhibit similar critical behavior.

In summary, in this work we have presented a different, bifurcation-based perspective
to the focusing problem in the ubiquitous NLS equation.
We have
``factored out'' the self-similarity group action through the MN
dynamics approach and set the problem up as a steady state problem
in the latter framework.
This allowed us to identify stable
self-focusing solutions for the original problem (which actually
are unstable for MN dynamics).
We were able continue the relevant
branches, constructing the supercritical pitchfork-like bifurcation diagram
that constitutes the focusing instability for this novel (mixed
Hamiltonian-dissipative) dynamical system.
We were able to identify
the eigenvalues of the focusing steady states of the MN dynamics and
to explain their spectrum on the basis of the knowledge of the
``soliton branch'' spectrum.
The theoretical study of the normal
form of the MN-dynamics that could provide analytical justification
of the scaling behavior near the transition is currently under study
and will be reported in future publications.

We thank D. Aronson, S. Betelu, K. Lust, B. Malomed, D. McLaughlin
 and C. Rowley
for stimulating discussions. The support of
of AFOSR and NSF (IGK) and of the University of Massachusetts (PGK)
is also gratefully acknowledged.

\newpage

\begin{references}
\bibitem{Taylor} This is a standard technique in finding travelling waves,
where the relevant symmetry that creates the degeneracy is translational
invariance. It was only recently, however,
appreciated that this can be applied to any one parameter family of
solutions induced by the action of a symmetry group.
\end{references}
\end{document}